\documentclass[twocolumn,showkeys,preprintnumbers,amsmath,amssymb,prb,aps]{revtex4}

\usepackage{graphicx}

\pdfoutput=1

\begin{document}

\title{Hierarchy of adhesion forces in patterns of photoreactive surface
layers}

\author{Gregor Hlawacek}
\author{Quan Shen}%
\author{Christian Teichert}
 \email{teichert@unileoben.ac.at}
 \homepage{http://www.unileoben.ac.at/~spmgroup}
\affiliation{%
Institute of Physics, University of Leoben, 8700 Leoben, Austria
}%

\author{Alexandra Lex}
\altaffiliation[Current address: ]{Institute of Physical Chemistry,
University of M\"unster, 48149 M\"unster, Germany}
\author{Gregor Trimmel}
\affiliation{
Institute for Chemistry and Technology of Materials, Graz University of
Technology, 8010 Graz, Austria
}%

\author{Wolfgang Kern}
\affiliation{
Institute of Chemistry of Polymeric Materials, University of Leoben, 8700 Leoben,
Austria}

\date{\today}

\begin{abstract} 
  
  Precise control of surface properties including electrical
  characteristics, wettability, and friction is a prerequisite for
  manufacturing modern organic electronic devices. The successful
  combination of bottom up approaches for aligning and orienting the
  molecules and top down techniques to structure the substrate on the nano
  and micrometer scale allows the cost efficient fabrication and
  integration of future organic light emitting diodes and organic thin film
  transistors. One possibility for the top down patterning of a surface is
  to utilize different surface free energies or wetting properties of a
  functional group. Here, we used friction force microscopy (FFM) to reveal
  chemical patterns inscribed by a photolithographic process
  into a photosensitive surface layer. FFM allowed the simultaneous
  visualization of at least three different chemical surface terminations.
  The underlying mechanism is related to changes in the chemical
  interaction between probe and film surface.

\end{abstract}

\keywords{Friction force microscopy, photoreactive thin layers,
photo-lithography, isomerization}
\maketitle

\section{Introduction}

Modern low cost devices are increasingly based on organic semiconductors.
This important class of materials allows to achieve well-priced thin film
transistors and optical components such as light emitting diodes. An important
intermediate step in this technology is the possibility to control the
growth behaviour of active organic materials in terms of orientation
and structure on short length scales. To do so patterned thin surface
layers or self-assembled monolayers can be used. The patterning of such
films can be achieved in various ways including but not limited to soft
lithography,\cite{Larsen1997,Xia1996} scanning probe techniques (dip pen
lithography,\cite{Piner1999} nanografting,\cite{Xu1999,Liu2000} etc.),
energetic beams (UV-light,\cite{Sugimura2001}
electrons,\cite{Carr1997,Lercel1993} etc.) and many more.\cite{Smith2004}

Here, we present a friction force microscopy (FFM) study of thin surface
layers of a photosensitive thiocyanate-functionalized trialkoxysilane on
silicon oxide (SiO$_x$). These films can be reliably prepared on this
technological important surface with a high degree of control over the
final film thickness. However, more important is the fact that these films
can easily be modified by UV light and subsequently
functionalized.\cite{Lex2008}

As different end groups of the molecule will have different interaction
with the probe of the atomic force microscope, FFM\cite{Meyer1990} allows
to differentiate between them. Using this approach, four different
terminations could be distinguished and hierarchically ordered on a sample
that has been subsequently irradiated twice using line masks with different
feature spacings. As a result the different terminations could be
hierarchically ordered by their interaction strength with the AFM
probe.

\section{Experimental}

For the preparative work of the organic thin surface layers, hazardous
chemicals and solvents are used (ammonium thiocyanate, methanol,
propylamine, 2,2,2-trifluoroethylamine, and piranha solution). In addition,
piranha solution is explosive, and its preparation is highly exothermic (up
to 120$^\circ$C). Therefore, reactions must be carried out in a fume hood,
and protective clothes and goggles must be used! UV irradiation causes
severe eye and skin burns. Precautions (UV protective goggles and gloves)
must be taken!

\begin{figure}[!tbp]
  \begin{center}
    \includegraphics[width=.45\textwidth]{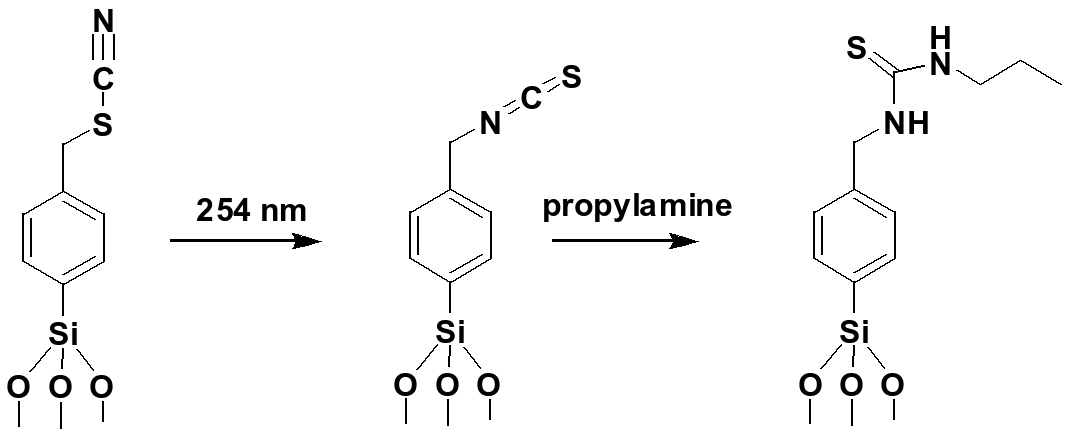}
  \end{center}
  \caption{Reaction pathway and products investigated by friction force
  microscopy.}
  \label{fig:reactions}
\end{figure}

The photoreactive surface layers were prepared by immersion of pretreated
(Piranha solution) boron doped silicon wafers into a solution of
trimethoxy[4-(thiocyanatomethyl)phenyl]silane (Si-SCN) in toluene.  X-ray
reflectivity measurements revealed a film thickness of 6\,nm for these
films. It has to be emphasized here, that obviously this is not a monolayer
but an oligolayer with a thickness corresponding to five or six individual
layers (assuming upright standing molecules). The formation of oligolayers
is attributed to cross linking of the trimethoxy-silane groups in the
presence of water.\cite{Wang2003} In a subsequent step, the samples were
illuminated by UV light under inert gas to avoid photo-oxidation (254\,nm,
80\,$\mathrm{{mJ}/{cm^2}}$). The illumination leads to an isomerization of
the benzyl thiocyanate (Si-SCN) group to the corresponding benzyl
isothiocyanate (Si-NCS). This illumination step was done by utilizing a
contact mask with equidistant lines and spaces to create a pattern on the
surface which consists of alternating stripes of Si-SCN and Si-NCS. For
selected samples an additional post-isomerization modification was
performed by exposing the surface to vapours of propylamine. In this
process the isothiocyanate group reacts to the corresponding thiourea group
(Si-PA). The sequence of the reaction steps together with the respective
molecular structure is shown in Fig.~\ref{fig:reactions}. A detailed
description of the film preparation has already been given
elsewhere.\cite{Lex2008}

The AFM results were obtained with a Digital Instruments Multimode IIIa
atomic force microscope. To
reduce damage to the film, the topographic images were recorded in
intermittant contact mode, eliminating effectively lateral forces between
the tip and the sample surface. For intermittant mode normal Si probes with a
typical resonance frequency of 300\,kHz were used. For roughness
characterization, the root mean square roughness $\sigma$, the lateral
correlation length $\xi$, and the Hurst parameter $\alpha$ were calculated
from the images using the height-height correlation function
(HHCF).\cite{Zhao1998} All roughness parameters have been obtained by
analyzing at least three independent 5\,$\mu$m images. FFM (also called
lateral force microscopy (LFM) or chemical force
microscopy)~\cite{Meyer1990,Mate1987} is a special type of contact mode
atomic force microscopy. Lateral forces acting on the tip will twist the
cantilever, when scanned perpendicular to its long axis, leading to a
deflection of the laser on the four quadrant photodiode in lateral
direction. The twist of the cantilever depends on the friction between the
tip and the sample surface. As cantilevers, specially designed FFM silicon
rectangular beam type cantilevers are used. The cantilevers have a nominal length
of 225\,$\mu$m and a force constant of typical 0.2\,N/m.

The lateral force acting on the tip is influenced by the friction
coefficient between tip and sample surface. This coefficient depends on the
interaction between the tip and the terminating group of the molecules
forming the thin film.\cite{Frisbie1994} For a clearer contrast, FFM
images are calculated from trace and retrace images obtained simultaneously
with the topographic image.\footnote{Gwyddion version 2.9
{(http://gwyddion.net/)} was used for image preparation and extraction of
roughness parameters.} This effectively reduces false FFM contrast
originating from the surface morphology. The presented images are therefore
always calculated from $(trace-retrace)/2$.  For the presented FFM images no scale is given as
no force calibration was performed prior to the measurement. Thus information that can
be obtained is purely qualitative, however, with sufficient accuracy
allowing to establish a hierarchy of adhesion forces.\looseness=1

\section{Results} 

\begin{figure*}[!tb]
  \includegraphics[width=\textwidth]{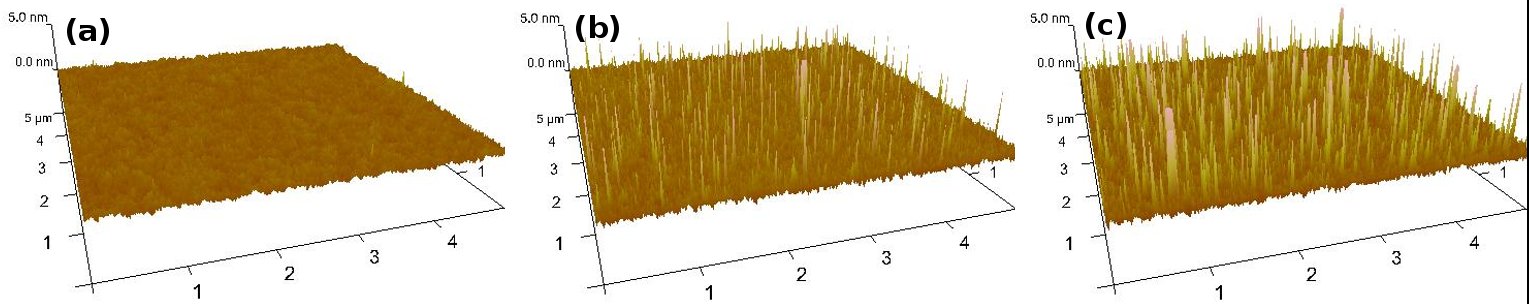}
  \caption{AFM topography images recorded in intermittant mode of (a) the
  SiO$_2$ substrate, (b) the film after Si-SCN deposition and subsequent
  flood illumination with 254\,nm UV light, and (c) modification with propylamine.
  (z-scale in all images is 5\,nm.)}
  \label{fig:unpatt}
\end{figure*}

Figure~\ref{fig:unpatt} presents AFM topography images demonstrating the
effect of film preparation on surface roughness. Homogeneous films of
Si-SCN (b), and propylamine modified Si-NCS (Si-PA) (c) films were prepared
and compared to the surface of the bare substrate (a).

The surface of the substrate (Fig.~\ref{fig:unpatt}(a)) shows a uniform
featureless topography as expected for a silicon dioxide surface. The root
mean square (rms) roughness of $\sigma$=0.2\,nm, the lateral correlation
length $\xi$=30\,nm, and the Hurst parameter $\alpha$=0.5 confirm the
qualitative observation. Other investigations report much larger
correlation length for SiO$_2$ measured by X-ray and optical
techniques.\cite{Teichert1995} The smaller values given here are related to
the shorter length scales that can be evaluated with AFM.
Deposition of a thin layer of Si-SCN and subsequent illumination with
254\,nm UV light for 20\,min (resulting in a Si-NCS film shown in
Fig.~\ref{fig:unpatt}(b)) does not lead to a strong change in the roughness
parameters: $\sigma$=0.3\,nm, $\xi$=30\,nm and $\alpha$=0.5. As 
mentioned above, X-ray reflectivity measurements revealed a film thickness
of 6\,nm for this layer.\cite{Lex2008}
Modifying the surface with vapors of propylamine (Fig.~\ref{fig:unpatt}(c))
leads to a further slight increase in surface roughness ($\sigma=0.4$\,nm
$\xi=20$\,nm, and $\alpha=0.5$).
The changes in the roughness parameters can be mainly attributed to contamination
of the sample surface due to handling and residue from the chemical
compounds used during modifications and cleaning.

\begin{figure}[!tbp] 
  \includegraphics[width=.45\textwidth]{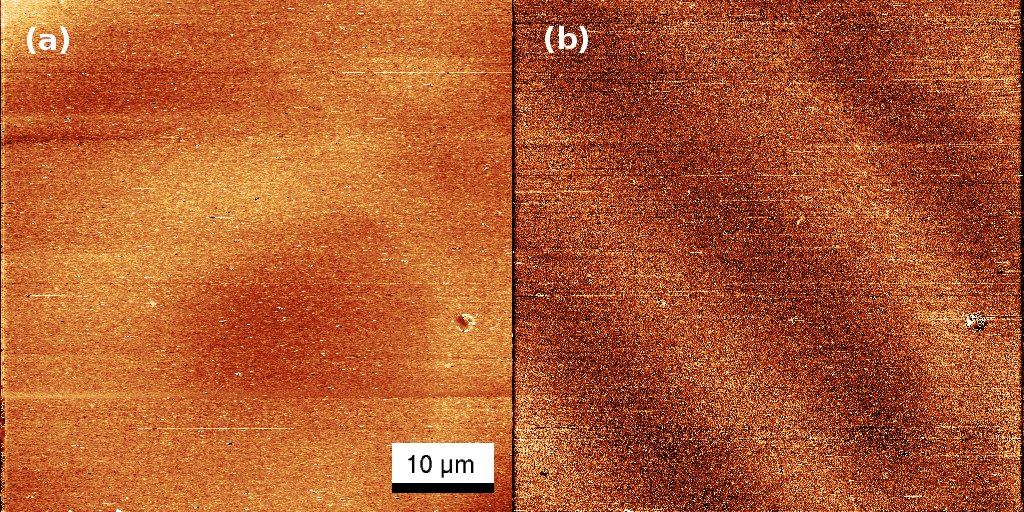}%
  \caption{Topographic (z-scale: 10\,nm) (a) and corresponding Friction
  Force image (b) from a Si-SCN film patterned with 254\,nm UV light
  through a 10\,$\mu m$ mask. In the FFM image (b) bright areas correspond
  to Si-NCS terminated areas showing high friction. The dark stripes are
  the non-illuminated Si-SCN areas.}
  \label{fig:scn-ncs}
\end{figure}

In the following, FFM data of the patterned films are presented.
Figure~\ref{fig:scn-ncs} compares the topographic image (a) obtained on
 a Si-SCN layer that was illuminated through a stripe mask with 10\,$\mu$m
lines and 10\,$\mu$m spaces to the simultaneously recorded FFM image (b).
Whereas in the topography image no stripe pattern is visible, a clear
stripe pattern with a 10\,$\mu$m pitch appears in the FFM image. In all
FFM images presented, bright areas mean higher friction and dark ones
correspond to lower friction. It will be demonstrated below that the high
friction areas correspond to the illuminated Si-NCS stripes and the low
friction areas are the pristine Si-SCN stripes covered by the mask during
illumination.

\begin{figure*}[!tbp]
  \includegraphics[width=\textwidth]{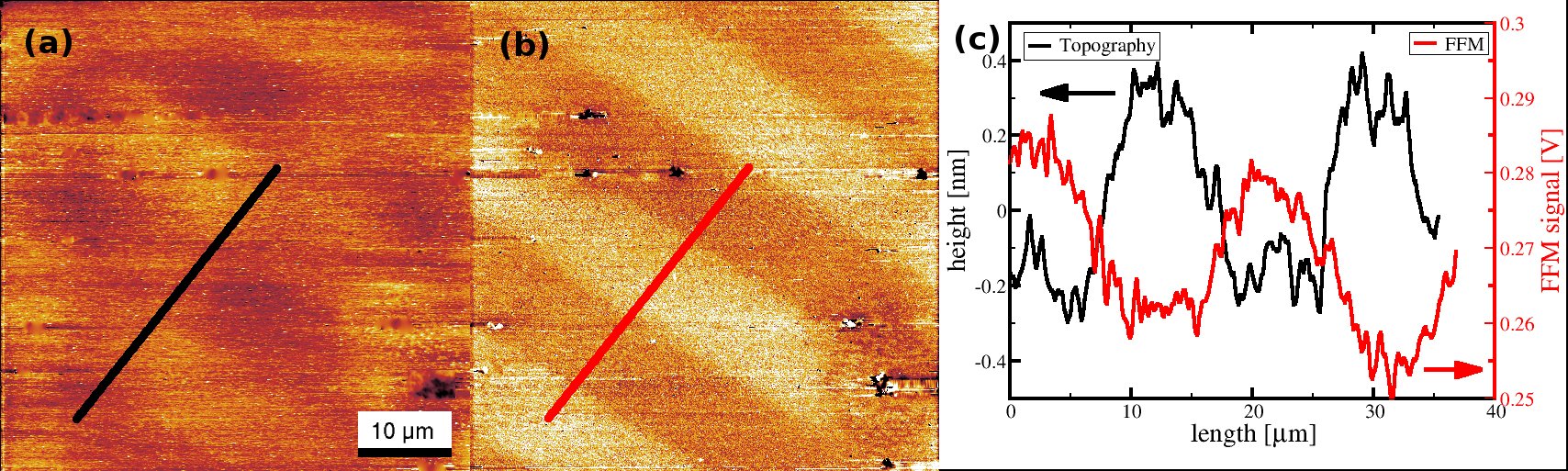}
  \caption{50\,$\mu m$ topographic (z-scale: 5\,nm) (a) and FFM (b) image
  from a patterned Si-SCN/Si-PA film. The pattering has been done through a
  contact mask with a 10\,$\mu m$ pitch. (c) Indicated cross sections
  reveal the expected 10\,$\mu m$ pitch in both topography and friction
  contrast. The Si-PA stripes are roughly 0.6\,nm higher than the Si-SCN
  stripes.}
  \label{fig:scn-pa}
\end{figure*}

Figure~\ref{fig:scn-pa} shows AFM images obtained after the patterned
sample described above has been exposed to vapours of propylamine. While
the topographic image presented in Fig.~\ref{fig:scn-pa}(a) shows a weak
stripe pattern with the expected spacing, the corresponding FFM image in
Fig.~\ref{fig:scn-pa}(b) allows a clear identification of the line pattern
created by contact lithography and of errors in the masking process (lower
right corner). From the combined cross section in Fig.~\ref{fig:scn-pa}(c)
the height difference of 0.5\,nm between the modified Si-PA and the
pristine Si-SCN stripes is clearly discernible. The addition of an
alkyl group to the molecule will result in an increase of the film
thickness. An increased
height is therefore only expected for the modified parts of the
surface, which are the Si-PA areas. These areas show a lower friction
signal. Areas of lower height correspond to non-illuminated (and unmodified)
zones of the layer containing Si-SCN structures, which give a higher
friction than the Si-PA structure. The increase in layer thickness after
reaction with propylamine was 0.6\,nm which equals 10\% of the initial layer
thickness. This corresponds to results obtained with photoreactive polymers
containing benzyl thiocyanate side groups.\cite{Kavc2002a}

\begin{figure*}[!tbp]
  \includegraphics[width=\textwidth]{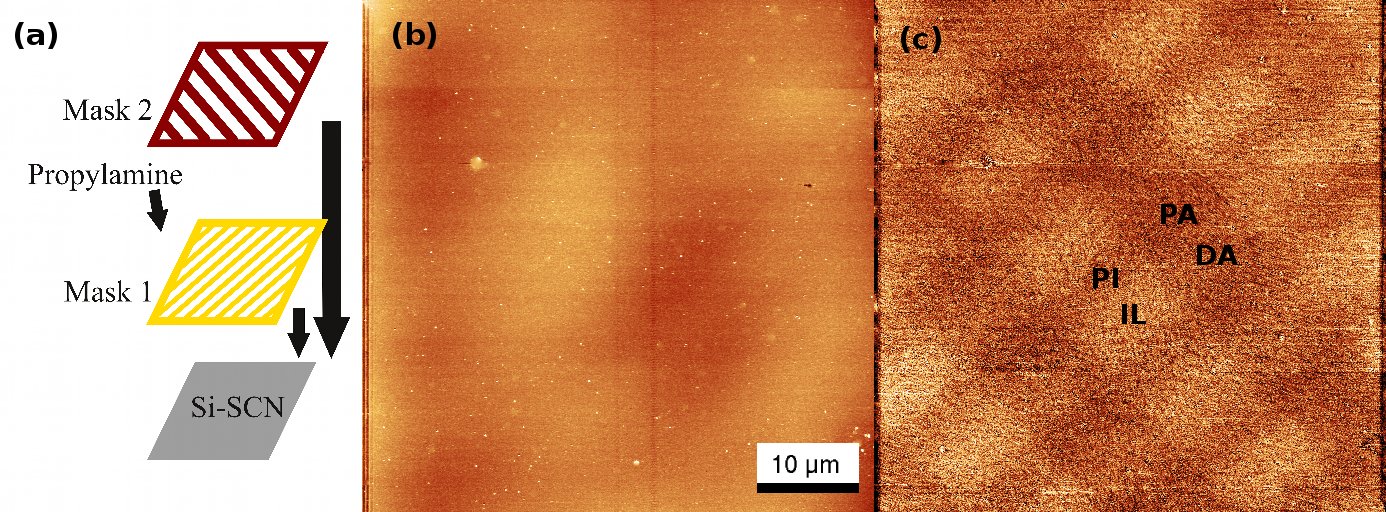}%
  \caption{(a) Scheme of sample preparation: Illumination of Si-SCN through
  a 5\,$\mathrm{\mu}$m mask (Mask 1); exposure to propylamine; second
  illumination through a 10\,$\mathrm{\mu}$m mask (Mask 2). (b) resulting
  topographic (z-scale: 10\,nm) and (c) corresponding FFM image from the
  double patterned Si-SCN/Si-PA/Si-NCS film. The respective materials are
  indicated in the FFM image (see text). 
  }
  \label{fig:cross}
\end{figure*}

In a further experiment, the surface has been exposed twice to illumination
using crossed masks with an intermediate propylamine reaction step. The
whole sample preparation process is sketched in Fig.~\ref{fig:cross}(a). In
a first step, a mask with 5\,$\mu$m lines and spaces was employed during UV
illumination. After the Si-SCN film has been exposed to vapors of
propylamine, a surface layer made up of alternating 5\,$\mu$m stripes of
Si-SCN and Si-PA is created. The resulting pattern is similar to the one presented in
Fig.~\ref{fig:scn-pa}. This modified film was now illuminated for a second time
through a mask with 10\,$\mu$m lines and spaces oriented perpendicular to
the first mask pattern.
During this step both stripes, Si-PA as well as Si-SCN, are illuminated
partly. It can be expected that the Si-PA surface will not change
significantly during this process. However, the remaining Si-SCN stripes
will be converted into alternating 10\,$\mu$m patches of Si-SCN and Si-NCS.
The resulting surface morphology and the friction image are presented in
Fig.~\ref{fig:cross}(b,c). As before, the main features in the topographic image
(Fig.~\ref{fig:cross}(b)) are not related to the mask process but result
from contamination and small long range undulations in the SiO$_x$ surface
of the wafer. However, the FFM image shows a clear pattern of regular 5 by
10\,$\mu$m patches of four different shadings, i.e. friction levels. We can
identify the four areas using the information on the friction contrast
obtained from the previous samples. The dark areas in Fig.~\ref{fig:cross}(b)
are the Si-PA areas (labeled PA) created in the first illumination step.
The neighboring brighter patches (PI) in the 5\,$\mu$m stripe are the Si-PA
areas that were subsequently illuminated a second time. In the neighboring 5\,$\mu$m
stripes (shadowed during the first illumination) one patch has been
protected by the mask in both illumination steps (DA) while the other one
has been exposed to UV-light a single time during the second illumination
(IL). The last two areas are comparable to those shown in
Fig.~\ref{fig:scn-ncs} and therefore allow the identification of the
stripes in Fig.~\ref{fig:scn-ncs}(b). The areas marked PA and DA are
comparable to the combination shown in Fig.~\ref{fig:scn-pa}.

The observed friction contrast can therefore be ordered in the following way:
The highest tip-film interaction and therefore the largest friction is
observed for Si-NCS (IL) followed by Si-SCN (DA) and the two propylamine
modified surfaces (PI and PA in Fig.~\ref{fig:cross}(b)).

A possible explanation fort his sequence can be given by the different
polarity of the individual endgroups at the surface (thiocyanate (Si-SCN),
isothiocyanate (Si-NCS), and propyl (CH$_2$-CH$_2$-CH$_3$, Si-PA)) and
by the stiffness of the molecular endgroups. 

Comparing  data on the dipole moment of ethyl isothiocyanate (3.67 Debye)
and ethyl thiocyanate (3.33 Debye) as well as on the surface tension
$\gamma$ (at 20$^\circ$C) of these compounds (ethyl isothiocyanate: $\gamma$ =
36.0 mN/m, and ethyl thiocyanate: $\gamma$ = 34.8 mN/m) it is found that
isothiocyanates are of higher polarity than the corresponding
thiocyanates.\cite{Dean1999}
Assuming that the friction force between the surface and the silicon tip
(which is covered with a native oxide layer) increases with the polarity of
the surface, for the UV illuminated regions (containing NCS units at the
surface) a higher friction force will be recorded than for the
non-illuminated regions bearing SCN units. 

After reaction of the photogenerated NCS groups with propylamine (PA), the
surface is terminated with non-polar alkyl groups. It is therefore not
surprising that the post-exposure derivatization with propylamine will
result in lower friction force. Also the stiffness of the
molecule is reported to influence the resulting friction coefficient.
\cite{Overney1992,Bhushan2001} The flexible alkyl group that terminates
Si-PA will therefore also reduce the observed friction by bending under the
applied normal load. In contrast, the shorter and stiffer thiocyanate and
iso-thiocyanate groups (Si-NCS has two double bonds
between  sulfur, carbon, and nitrogen; Si-SCN has a triple bond between
carbon and nitrogen while the sulfur is linked by two single bonds) can not
bend under the applied load.

The intermediate friction contrast observed for areas that were illuminated
and modified with propylamine, and then illuminated for a second time, can
also be explained that way. During the second illumination, residual SCN
units -- which have remained unreacted in the first illumination step -- are
converted into NCS groups leading to an increased interaction between
the surface and the AFM tip. NCS groups which have reacted with propylamine (to
yield propyl thiourea units) are expected to remain unchanged during the
second illumination step since N,N'-dialkyl substituted thiourea groups are
stable under UV light. Especially from the last sample, the hierarchy in the
interaction forces (Si-NCS $>$ Si-SCN $>$ Si-PA) between the individual
terminating groups becomes evident.

\section{Conclusion}
\label{sec:con}

For lithographically patterned
photoreactive surface layers we demonstrated that friction force microscopy is not only able to
distinguish between different head groups, but also that this can be done simultaneously
for at least three different terminations. The method is able to detect
minute changes in the molecular structure. The case of the
thiocyanate (Si-SCN) and isothiocyanate (Si-NCS) head groups is especially
relevant since these are educts and products of an isomerization reaction.
FFM is able to discern the two terminations, although, only the
sequence of the terminating three atoms is altered. However, the resulting change in
polarity of the terminating molecular groups is large enough to be detected
by FFM as different friction levels. In addition, the post
isomerization modification with propylamine could be clearly visualized
with FFM. In this case, a further reduction of polarity together with a
change in the stiffness of the molecule leads to the lowest friction observed
in the investigated system.

Currently, contact angle measurements are underway to obtain an independent
confirmation of the observed hierarchy. The next step in future work will
be to quantify the adhesive and frictional forces responsible for the
qualitative results presented here. For this task a well known approach
suggested in literature will be used to calibrate the AFM probes with
sufficient accuracy.\cite{Varenberg2003} These studies shall include the use
of functionalized tips to fine tune the probe sample
interaction.\cite{Noy1995}

\acknowledgements
  This work was supported by Austrian Science Fund FWF projects S9707-N08
  and S9702-N08. 


\end{document}